\documentclass[jap,graphicx, floatfix]{revtex4-1} % for checking your page length
\usepackage{amsfonts, amsmath, graphicx}
\usepackage[version=3]{mhchem}

\begin{document}
\title{Adiabatic approximation for a uniform DC electric field} 
\author{Zhuo Bin Siu}
\affiliation{Computational Nanoelectronics and Nanodevices Laboratory, Electrical and Computer Engineering Department, National University of Singapore, Singapore} 
\author{Mansoor B. A. Jalil} 
\affiliation{Computational Nanoelectronics and Nanodevices Laboratory, Electrical and Computer Engineering Department, National University of Singapore, Singapore} 
\author{Seng Ghee Tan} 
\affiliation{Data Storage Institute, Agency for Science, Technology and Research (A*STAR), Singapore} 

\begin{abstract}
	In this work, we show that the disorder-free Kubo formula for the non-equilibrium value of an observable due to a DC electric field, represented by $E_x\hat{x}$ in the Hamiltonian,  can be interpreted as the standard time-independent theory response of the observable due to a time- and position-\textit{independent} perturbation $H_{MF}$. We derive the explicit expression for $H_{MF}$ and show that it originates from the adiabatic approximation to $\langle k|E_x\hat{x}$ in which transitions between the different eigenspinor states of a system are forbidden. The expression for $H_{MF}$ is generalized beyond the real spin degree of freedom to include other spin-like discrete degrees of freedom (e.g. valley and pseudospin).  By direct comparison between Kubo formula and the time-independent perturbation theory, as well as the Sundaram-Niu wavepacket formalism, we show that $H_{MF}$ reproduces the effect of the E-field, i.e. $E_x\hat{x}$, up to the first order. This replacement suggests the emergence of a new spin current term that is not captured by the standard Kubo formula spin current calculation. We illustrate this via the exemplary spin current for the heavy hole spin 3/2 Luttinger system. Finally, we apply the formalism and derive an analogous $H_{MF}$ for the effects of a weakly position-dependent coupling to the spin-like internal degrees of freedom. This gives rise to an anomalous velocity as well as spin accumulation terms in spin$\otimes$pseudospin space in addition to those contained explicitly in the unperturbed Hamiltonian. 
\end{abstract} 
\maketitle

\section{Introduction}

	In the Spin Hall Effect (SHE) \cite{SHE1,SHE2,SHE3, SHE4, SHE5}, the passage of an in-plane electric field in a two-dimensional electron gas (2DEG) with spin orbit coupling (SOC) leads to the emergence of an out-of-plane spin accumulation. 	Murakami \cite{Murakami} and Fujita \cite{Grp1,Grp2,Grp3,Grp4,SGTSciRep}, and their respective coauthors, had independently studied the SHE. They showed that the out-of-plane spin accumulation can be understood as the response of the charge carriers as their spins align adiabatically with the momentum-dependent SOC field. The direction of the SOC field changes in time due to the change in the momentum of the charge carriers as they accelerate under the electric field.  Mathematically, the electric field gives rise to an effective out-of-plane magnetization term perpendicular to the SOI field in the Hamiltonian. 
	
	The usual derivation \cite{Grp1,Grp2,Grp3,Grp4,SGTSciRep} of this effective magnetization involves a unitary transformation from the laboratory frame, where the spin quantization axis is conventionally taken to be an arbitrary \textit{fixed} $z$ axis, to the eigenbasis frame where the spin quantization axis now points along the SOC field. For concreteness, consider the spin 1/2 Hamiltonian 
\begin{equation}
	H(\vec{k}) = B_0(\vec{k})\mathbf{I}_\sigma + B(\vec{k})\cdot\vec{\sigma} + E_x\hat{x} \label{H0} 
\end{equation} 
where $B_0(\vec{k})\mathbf{I}_\sigma$ are the spin-independent terms, $B(\vec{k})\cdot\vec{\sigma}$ is a momentum dependent SOC field that may also possibly have momentum independent components (for example, a Rashba SOC with a uniform $x$ magnetization $M_x$ would give $\vec{B}(\vec{k}) = (\alpha k_y + M_x, -\alpha k_x)$), and $E_x \hat{x}$ represents an electric field in the $x$ direction.  We denote the $m$th eigenspinor of $\vec{B}(\vec{k})\cdot\vec{\sigma}$ as $|\chi_m(\vec{k})\rangle$. 

In the unitary transformation formalism, we consider a unitary transformation $U$ which diagonalizes $\vec{B}(\vec{k})\cdot\vec{\sigma}$ so that $U B(\vec{k})\cdot\vec{\sigma} U^\dagger = |\vec{B}|\sigma_z$. Now since $U$ is momentum dependent due to the momentum dependence of $\vec{B}(\vec{k})$, we have
\begin{equation}
	UH(\vec{k})U^\dagger = B_0(\vec{k})\mathbf{I}_\sigma + |\vec{B}(\vec{k})|\sigma_z + E_x ( \hat{x} + i U\partial_{k_x}U^\dagger) \label{UHU} 
\end{equation} 
where there is now an `additional' $i U\partial_{k_x}U^\dagger$ term due to the non-commutativity between the momentum and position operators. $i U\partial_{k_x}U^\dagger$ has diagonal spin components, which point along the same spin direction as $\vec{B}$, as well as \textit{off-diagonal} components which point in a spin space direction perpendicular to $\vec{B}$. The latter is usually identified as the source of the out-of-plane spin accumulation in the SHE when $\vec{B}$ is the in-plane SOC field. This, however, raises the issue of how we can reconcile the off-diagonal elements of $i U\partial_{k_x}U^\dagger$ with employing an adiabatic approximation. The off-diagonal elements of $i U\partial_{k_x}U^\dagger$ correspond to transitions between eigenstates of $\vec{B}\cdot\vec{\sigma}$ whereas the adiabatic approximation is usually associated with forbidding transitions between eigenstates. One of the two main motivations of this work, then, is to clarify how this unitary transformation formalism is actually consistent with taking an adiabatic approximation. 

Besides the emergence of spin accumulation, another hallmark of the SHE is the appearance of currents of out-of-plane spin flowing in the in-plane direction perpendicular to the applied electric field. One common method to obtain these spin currents is by using the Kubo formula \cite{Sinova} which, in the absence of impurity scattering, gives the non-equilibrium value of an observable $O$ due to a uniform DC electric field in the $i$th direction as 
\begin{equation}
	\delta O / E_i  = \sum_{\vec{k}, m' \neq m} (n_{m',\vec{k}} - n_{m, \vec{k}})\frac{ \mathrm{Im} \langle \chi_{m'}(\vec{k})|O|\chi_{m}(\vec{k})\rangle\langle \chi_m(\vec{k})|j_i|\chi_{m'}(\vec{k})\rangle }{(e_{m,\vec{k}} - e_{m', \vec{k}})^2}. \label{Kubo} 
\end{equation} 
where $n_{m, \vec{k}}$ and $e_{m,k}$ are  the Fermi-Dirac occupancy and energy of the $|\chi_m(\vec{k})\rangle$ state respectively. 
Not surprisingly, putting $O = \sigma_z$ gives consistent results for the out-of-plane spin accumulation given by the unitary transformation formalism described earlier. Moreover, some of us ( S. G. T. and M. B. A. J.) have pointed out in our earlier works \cite{Grp2, Grp4, SGTSciRep} that the Berry curvature anomalous velocity in the $i$th direction $\propto  (\partial_{k_i}\hat{b}\times\partial_{k_j}\hat{b})\cdot\hat{b} E_j$, $\hat{b} \equiv \vec{B}/|\vec{B}|$ that is more commonly derived by putting $O=v_i$ into the Kubo formula Eq. \ref{Kubo}, or by applying the Sundaram-Niu wavepacket formalism \cite{SN0}  can also be obtained from the Heisenberg equation of motion for $\dot{x}_i = -i[\hat{x}_i, \tilde{H}]$ by including a term proportional to the off-diagonal terms of $i U\partial_{k_i}U^\dagger$ into $\tilde{H}$. The second motivation of this paper is therefore to clarify the link between the unitary transformation formalism and the Kubo formula. 

We claim that to first order in perturbation theory, spin accumulation and velocities can be predicted by replacing $E_x\hat{x}$ in Eq. \ref{H0} by $H_{MF} \equiv E_x \sum_{m,m' \neq m} |\chi_{m'}(\vec{k} )\rangle i\langle \partial_{k_x} \chi_{m'}(\vec{k} )|\chi_m(\vec{k} )\rangle \langle \chi_m(\vec{k})|$.  (The subscript $MF$ stands for Murkami-Fujita.) After introducing the notation we shall be using in this paper, we first motivate the introduction of $H_{MF}$ by showing that the  Kubo formula Eq. \ref{Kubo}  can be interpreted  as the first-order expectation value of the observable under the perturbation of $H_{MF}$. We next show that $H_{MF}$ originates from taking the adiabatic approximation of retaining only the \textit{diagonal} elements in the eigenspinor basis representation of the position operator $U\hat{x}U^\dagger$ in Eq. \ref{UHU}. This discarding of the off-diagonal terms physically corresponds to preventing transitions between different eigenspinor states.  We then show that the results of replacing $E_x\hat{x}$ with $H_{MF}$ in the Hamiltonian and applying the Heisenberg equation of motion reproduces the same results as the Sundaram-Niu wavepacket formalism \cite{SN0} for the spin evolution and charge current.  The wavapacket formalism provides a physical justification for discarding $E_x\hat{x}$ in  our effective Hamiltonian. 

The addition of $H_{MF}$ into the effective Hamiltonian suggests the consideration of a new contribution for the spin  $\alpha$ current flowing in the $j$th direction of the $|\chi_m(\vec{k})\rangle$ state $\frac{1}{2} \langle \chi_m(k)| \{ \sigma_\alpha, -i[x_j, H_{MF}]  \} |\chi_m(\vec{k})\rangle$ that is \textit{not} captured by the usual practice of putting $O = \frac{1}{2} \{ \sigma_\alpha, -i[x_j, B_0(\vec{k})\mathbf{I}_\sigma + \vec{B}(\vec{k})\cdot\vec{\sigma}] \}$  into the Kubo formula Eq. \ref{Kubo}.  We evaluate this spin current for the Luttinger spin 3/2 system. 

Using the same formalism we used to show the origin of $H_{MF}$ due to an electric field $E_x\hat{x}$, we next derive the analogous $H_{MF; \text{mag}}$ for the perturbation due to a linear variation of the magnetization $(\partial_x \vec{M})\cdot\vec{\sigma}\hat{x}$ . We show that the resulting $H_{MF;\text{mag}}$ also results in a spin-dependent anomalous velocity and spin accumulations. 

\section{System definition}
\label{sec:sysDef} 
Here we study translation-invariant systems with momentum-dependent spin-orbit interactions and possibly other discrete internal degrees of freedom perturbed by a uniform DC electric field. For example, many emerging material systems of interest in spintronics like silicene \cite{Sil1,Sil2,Sil3,Sil4} and  \ce{MoS2} \cite{Mo1, Mo2, Mo3}, possess discrete degrees of freedom such as the pseudospin and / or valley degrees of freedom, in addition to their real spins.  Another example of a system with a discrete internal degree of freedom is a topological insulator (TI) thin film which, unlike a semi-infinite thick TI slab, possesses both a top as well as a bottom surface where the surface states localized at different surfaces can couple to one another due to the finite thickness of the film  \cite{PRB80_205401,PRB81_041307,PRB81_115407}. The low energy effective Hamiltonian for the surface states of a TI thin film can thus be written as
\begin{equation}
	H = v(\vec{k}\times\vec{\sigma})\cdot\hat{z} \tau_z + \lambda \tau_x + \vec{M}\cdot\vec{\sigma} + E_z\tau_z \label{TIham1}
\end{equation}
where the $\sigma_i$s and $\tau_i$s are Pauli matrices. Besides the real spin of the charge carriers, denoted as $\vec{\sigma}$,  there is another discrete degree of freedom $\vec{\tau}$ associated with whether the charge carriers are localized nearer the upper (  $|+\tau_z\rangle\langle +\tau_z|$ ) or lower ($|-\tau_z \rangle\langle -\tau_z|$) surface of the film. The $\tau_x$ term then represents the coupling between the two surfaces of the film due to the finite thickness, and the $E_z\tau_z$ term the spin-independent potential energy difference between the top and bottom surfaces.  The $\vec{M}\cdot\vec{\sigma}$ term represents the exchange coupling to either ferromagnetic dopants or an adjoining ferromagnetic film.   

For simplicity, we collectively refer to all of the discrete internal degrees of freedom other than the real spin as `pseduospin', and denote the corresponding operators as $\tau$. We enumerate all the possible combinations of the $\sigma_i \otimes \tau_j$ operators (including $\mathbf{I}_\kappa\equiv \mathbf{I}_\sigma\otimes\mathbf{I}_\tau$) as $\kappa_k$. For instance, for the TI thin film Hamiltonian Eq. \ref{TIham1} we may define $\kappa_i \equiv \mathbf{I}_\sigma\otimes\tau_i$ for $i=1,2,3$, $\kappa_4 \equiv \mathbf{I}_\sigma\otimes\mathbf{I}_\tau$, $\kappa_{i} \equiv \sigma_x\otimes\tau_{i-4}$ for $i=5,6,7$, etc. 

We thus consider systems of the form 
\begin{eqnarray*} 
	H_0 &=& B_i(\hat{\vec{p}})\hat{\kappa}_i  \\
	H &=& H_0 + E\hat{x}.
\end{eqnarray*}
where we have placed hats on top of $\hat{\vec{p}}$ and $\hat{x}$ to emphasise that these are operators. In the absence of the electric field, the translation invariance of the system leads to the momentum being a good quantum number. For notational simplicity we temporarily restrict ourselves to working in one dimension and drop the vector arrow on top of $k$. (The extension to multiple spatial dimensions is trivial)  

Since momentum is a good quantum number, it is common to consider
\begin{equation}
	 (\langle k|\otimes \mathbf{I}_\kappa)  H_0(\hat{p}) = B_i(k)(\langle k|\otimes\kappa_i) \label{Hk1} 
\end{equation}
and call $H_0(k) \equiv B_i(k)\sigma_i$ \textit{the} Hamiltonian instead, for example as in Eq. \ref{H0}. Notice that the \textit{operator} $\hat{p}$ has been demoted to the \textit{numerical} eigenvalue $k$, and the $\langle k|$ on the right hand side is usually not written out but implied implicitly. The formal mathematical definition of $H_0(k)$ will however turn out to be important when we consider the expansion of the $E_x\hat{x}$ perturbation in the eigenspinor basis later in Sect. \ref{sec:HmfOrg}.

For a given momentum $k$, we denote the eigenstates of $H_0$ as $|k,m\rangle \equiv |k\rangle \otimes |\chi_m(k)\rangle$ where $|k\rangle$ is the $k$ momentum state ket, and $|\chi_m(k)\rangle$ is the $m$th eigenspinor of $H_0$ at momentum $k$. The corresponding eigenenergy $e_{k,m}$ satisfies $H_0(\hat{p}) |k,m\rangle = |k,m\rangle e_{k,m}$.

\section{Kubo formula} 
We briefly run through the textbook derivation of the Kubo formula Eq. \ref{Kubo}. The purpose of this overview is two-fold. We first wish to bring across the fact that the derivation is somewhat mathematically involved and not very physically transparent. We will show in this, and the next section, that Eq. \ref{Kubo} can be interpreted more transparently as giving the expectation value of $O$ due to the perturbation by the adiabatic approximation of $E_x\hat{x}$. We secondly also want to highlight one aspect of the derivation that is not very commonly discussed but which justifies our eventual replacement of $E_x\hat{x}$ with $H_{MF}$.  

The textbook derivation of the Kubo formula starts from linear response theory. Linear response theory tells us that the non-equilibrium value of the expectation value of an observable $O$ due to a uniform DC electric field of  amplitude $E_b$ in the $b$th direction is given by 
\begin{eqnarray*}
	\delta O &=& \lim_{\omega\rightarrow 0} E_b \frac{i}{\omega} C_{O j^b}(\omega)  \\
	&=& E_b\lim_{\omega\rightarrow 0}  \Big(  \frac{i}{\omega}  \frac{1}{\beta} \sum_{\omega_n} \mathrm{Tr} (O \mathcal{G}(i (\omega_n + q)j^b\mathcal{G}(i \omega_n)) \Big|_{q \rightarrow \omega + i\eta} \Big)
	\label{CrOJ} 
\end{eqnarray*}
where $C_{Oj^b}$ is the retarded correlation function between the observable $O$ and the charge current $j$ in the $b$th direction,  the $\mathcal{G}$s are complex frequency Matsubara Green's functions, and the sum $\omega_n$ occurs over Matsubara frequencies. The limit of taking $\omega \rightarrow 0$ corresponds to first assuming that the electric field is actually AC with frequency $\omega$ and then taking the DC limit of the electric field.

Using the standard summation formula $\frac{1}{\beta}\sum_{i\omega_n} n(i \omega_n)\exp(i\omega_n \tau) = \mp \sum_j \mathrm{Res}[f(z_j)] n(z_j)\exp(z_j\tau)$, where $z_j$ are the complex poles of the Fermi-Dirac distribution function $n$, to evaluate the complex frequency summation lands us at 
\begin{equation}
		\delta O = \sum_{\alpha,\beta}  \lim_{\omega \rightarrow 0} \frac{i}{\omega} \left( \frac{ n_\alpha-n_\beta}{e_\alpha-e_\beta + \omega + i \eta } \right) O_{\alpha,\beta}j^b_{\beta,\alpha} E_b. \label{KuboW1} 
\end{equation}
where $\alpha$ and $\beta$ are collective labels for the quantum numbers $(\vec{k},m)$ of the eigenstates of $H_0$ (which does not include the $E_x\hat{x}$), $|\vec{k},m\rangle$, and $O_{\alpha,\beta}\equiv \langle \alpha|O|\beta\rangle$.  The $\omega \rightarrow 0$ limit can be evaluated by exploiting the fact that 
\[
	\frac{1}{\omega}\frac{1}{\omega + x} = \frac{1}{x} (\frac{1}{\omega} - \frac{1}{\omega + x})	
\]
to split the summand into a divergent part which we discard, and a non-divergent part which we retain. We will have a bit more to say about the discarding of the divergent part later on. The non-divergent part in Eq. \ref{KuboW1} is 
\begin{equation}
	-i\frac{ n_\alpha - n_\beta }{ (e_\alpha-e_\beta )(\omega + i \eta + e_\alpha-e_\beta) } O_{\alpha,\beta}j^b_{\beta,\alpha}E_b \label{kuboX1} 
\end{equation}
where it is now safe to put $\omega \rightarrow 0$ explicitly. This gives the standard Kubo formula
\begin{equation}
	\delta O =  \sum_{\alpha \neq \beta} \frac{ n_\alpha-n_\beta}{(e_\alpha-e_\beta )^2}\mathrm{Im}\Big( O_{\alpha,\beta}j^b_{\beta,\alpha} \Big)  E_b.\label{TKNNkubo}
\end{equation}

In order to relate this to our results later and considering just the one-dimensional case for notational simplicity, we use the fact $\langle \chi_m|j|\chi_{m'} \rangle = \partial_k e_{k,m}\delta_{mm'} + \langle \partial_k \chi_m|\chi_{m'} \rangle (e_m-e_{m'})$ to obtain
\[
	\delta O = \sum_k \sum_{m \neq m'} \frac{ n(e_{k,m})-n(e_{k,m'}) }{ e_{k,m}-e_{k,m'} } \mathrm{Im} (\langle \chi_m(k)|O|\chi_{m'}(k) \rangle\langle \partial_k \chi_{m'}(k)|\chi_m (k))\rangle E.
\]

To isolate the contribution of a specific $|k,n \rangle$ state to $\delta O$, $O_{k,n}$ we can set the occupancy factor for this state to 1 and all other states to 0. Note that there are actually two sets of terms in the sums above which contribute -- one where $m=n, m' \neq n$ and $m\neq n, m' = n$. The contributions of these two sets are equal. We thus have  
\begin{eqnarray}
	\delta O_{k,n} &=& \sum_{m' \neq n} \frac{2}{e_{k,m'}-e_{k,n} } \mathrm{Im} (\langle \chi_n(k)|O|\chi_{m'}(k) \rangle\langle \partial_k \chi_{m'}(k)|\chi_n (k)\rangle \nonumber  E\\ 
	&=& 2\sum_{m' \neq n} \mathrm{Re} \langle \chi_n(k)|O| \frac{ |\chi_{m'}(k)\rangle i\langle \partial_k \chi_{m'}(k)|\chi_n(k)\rangle}{e_{k,n}-e_{k,m'}} E. \label{stdKub1} 
\end{eqnarray}

The derivation above is not very intuitive and is mathematically complicated with the use of complex frequencies and first assuming an AC electric field and then taking the limit of the electric field frquency going to 0. However Eq. \ref{stdKub1} admits a more physically intuitive interpretation.  In standard time-independent perturbation theory the first order correction to the state $|\chi_m \rangle$ due to a perturbation $V$ , which we denote as $|\chi_m^1\rangle$, is given by 
\begin{equation}
	|\chi_m^1\rangle = \sum_{m' \neq m} \frac{|\chi_{m'}\rangle\langle \chi_{m'}|V|\chi_m\rangle}{e_{m'}-e_m}. \label{chi1}
\end{equation}
Denoting the exact eigenspinor of $H_0(k)+V$ as ${|\tilde{\chi}_m\rangle}$, the expectation value of $O$ is 
\begin{eqnarray}
	\langle \tilde{\chi}_m |O|\tilde{\chi}_m\rangle &\approx& \langle \chi_m|O|\chi_m\rangle + 2\mathrm{Re}(\langle \chi_m|O|\chi_m^1\rangle) \nonumber \\
	&=& \langle \chi_m|O|\chi_m\rangle + 2\mathrm{Re}\left(\langle \chi_m|O| \sum_{m' \neq m}\frac{ |\chi_{m'}\rangle\langle \chi_{m'}|V|\chi_m\rangle}{e_{k,m}-e_{k',m'}}\right)
	 \label{Eo} 
\end{eqnarray}  to first order in $V$. Comparing Eq. \ref{Eo} to Eq. \ref{stdKub1}, it is evident that the corresponding $V$ in Eq. \ref{stdKub1} would be 
\begin{equation}
	V =  H_{MF} \equiv E_x \sum_{n,n' \neq n} |\chi_n'(k)\rangle i\langle \partial_k \chi_{n'}(k)|\chi_n(k)\rangle \langle \chi_n(k)| \label{Hmf} 
\end{equation}

This suggests that the effects of the $E_x\hat{x}$ term in the Hamiltonian on those observables which can be obtained by the Kubo formula can be reproduced, to first order in $E_x$, by replacing  $E_x\hat{x}$ with $H_{MF}$. This invites the question of where $H_{MF}$ comes from. We address this in the next section before showing that replacing  $E_x\hat{x}$ with $H_{MF}$ reproduces the current and spin dynamics predicted by the Sundaram-Niu formalism. 

\section{Origin of $H_{MF}$} 
\label{sec:HmfOrg} 

We pointed out at the end of Sect. \ref{sec:sysDef} that Hamiltonian, $H_0(k)$, actually corresponds to $\langle k |\otimes\mathbf{I}_\kappa H_0(\hat{p})$.  The addition of the perturbation $E\hat{x}$ presents the question of what form the corresponding $\langle k|\otimes\mathbf{I}_\kappa \hat{x}$ will take. 

To answer this question, we note that the spin$\otimes$pseudospin identity $\mathbf{I}_\kappa$ can be resolved in any set of orthogonal basis states. For concreteness, let us take the example that the $\mathbf{I}_\kappa$s correspond to spin 1/2 operators. One set of basis states that can be chosen is the spin up and down states, the `up' and `down' defined with respect to a constant spin quantization axis. The basis states are hence momentum independent. We shall, for brevity, refer to this set of basis states as the `laboratory frame basis' and where we need to refer to these states explicitly, always write the indices as $\lambda_i$ ($\lambda$ for \textit{l}aboratory)  with a subscript $i$ to differentiate between different basis states.  Another obvious choice of basis states to resolve $\mathbf{I}_\kappa$ into is in terms of the eigenspinors of $H_0(k)$, $|\chi_m(k)\rangle$, which unlike the laboratory frame basis, are momentum dependent. We show in the appendix that 
\begin{equation}
	\langle k|\hat{x} =  \sum_{m,m'}|\chi_m(k)\rangle\langle \chi_m(k)|i \partial_k \chi_{m'}\rangle\langle k, m'| +\sum_m |\chi_m(k)\rangle i\partial_k\langle  k,m| \label{kX} 
\end{equation}

We note that if we had expanded the spin identity operator in the momentum-independent laboratory basis $|\lambda_i\rangle$s instead, Eq. \ref{kX} reduces to the more familiar result
\begin{equation}
	\langle k|\hat{x} =  \mathbf{I}_\sigma (i\partial_k \langle k|) \label{kX1}
\end{equation} 

There is a bit of technical subtlety involved in relating Eq. \ref{kX} to the corresponding $(\hat{x}+iU\partial_{k_x}U^\dagger)$ term in Eq. \ref{UHU} where the ket-bra pairs in the operators are not written out explicitly.  There, one needs to be mindful of whether the resulting expressions are in the laboratory frame or in the eigenbasis frame. In Eq. \ref{UHU}, we made use of
\begin{equation}
	U\hat{x}U^\dagger = \hat{x} + iU\tilde{\partial}_k U^\dagger. \label{UXU} 
\end{equation} 
in which the unitary transformation on the left hand side of the equal sign brings the $\hat{x}$ operator in the \textit{laboratory} frame to the eigenspinor frame on the right hand side. Since this is a just a basis transformation $U$ is actually just an identity matrix and we read off that  $U^\dagger = \sum_{m,i} |\lambda_i\rangle\langle\lambda_i|\chi_m(k)\rangle\langle\chi_m(k)|$. The tilde that appears on the $\tilde{\partial_k}$ in Eq. \ref{UXU} reflects the subtlety that the $\partial_k$ within $U\partial_k U^\dagger$ in Eq. \ref{UXU} is understood to act only on the \textit{matrix elements} $\langle \lambda_i|\chi_m(k)\rangle$ and \textit{not} on the outermost ket-bra pair $|\lambda_i\rangle\langle \chi_m(k)|$ sandwiching the matrix elements. That is, $i\tilde{\partial_k}U^\dagger \equiv \sum_{m,i} |\lambda_i\rangle \partial_k(\langle\lambda_i|\chi_m(k)\rangle) \langle\chi_m(k)|$ -- the rightmost $\langle\chi_m(k)|$ bra is not differentiated.  This comes from the convention that in practical calculations the $U$ are written as numerical matrices, i.e. $U^\dagger = \begin{pmatrix} \langle \uparrow|+\rangle & \langle \uparrow|-\rangle \\  \langle \downarrow|+\rangle & \langle \downarrow|-\rangle \end{pmatrix}$ where $\pm$ are the eigenspinor states and $\uparrow/\downarrow$ are the lab spin up / down states, so that the only numbers that appear explicitly to be differentiated are the matrix elements. 

The correspondence of Eq. \ref{kX1} with Eq. \ref{kX} is that the second term on the right of Eq. \ref{kX} can be further expanded (by adding resolutions of identity) to yield 
\begin{equation} 
	\sum_m \big( |\chi_m(k)\rangle i\partial_k\langle k,m| \big) =  i \big( ( \partial_k \langle k|)\otimes\mathbf{I}_\sigma - \langle k|\otimes \sum_{m, m'} |\chi_m\rangle\langle\chi_m|\partial_k \chi_m'\rangle\langle \chi_{m'}| \label{cmkm}
\end{equation}
Adding this to the remaining $\sum_{m} i|\chi_m(k)\rangle\langle\chi_m(k)|\partial_k \chi_{m'}\rangle\langle k,m'| = i |\partial_k \chi_m'\rangle\langle k,m'|$ term in Eq. \ref{kX} leads to the recovery of Eq. \ref{kX1}. As a consequence Eq. \ref{UXU} should really be understood as 
\begin{eqnarray*}
	\hat{x}_l(k) &=& \hat{x}_e(k) + iU\tilde{\partial}_k U^\dagger \\
	|k\rangle \langle i\partial_k k|\otimes \mathbf{I}_\sigma &=& \sum_{m} |k,m\rangle  i\partial_k \langle k,m| + i \sum_{m,m'} |k,m\rangle \langle k| \otimes \langle \chi_m|\partial_k \chi_{m'}\rangle\langle\chi_{m'}|  \\
	&=& \sum_{m} |k,m\rangle  i\partial_k \langle k,m| - i |k,m\rangle \langle k |\otimes\langle \partial_k \chi_m|
\end{eqnarray*}
where we defined $\hat{x}_l(k)$ and $\hat{x}_e(k)$ in the second line (with the subscripts $l$ and $e$ for \textit{l}ab and \textit{e}igenspinor frames respectively, and the $k$ argument stressing that this is actually the $|k\rangle\langle k|$ projection of the position operator ). Notice that position operator $\hat{x}$ in Eq. \ref{UXU} actually has different definitions in the laboratory ($\hat{x}_l(k)$) and eigenbasis ($\hat{x}_e(k)$) frames. Compared to $\hat{x}_l(k)$, $\hat{x}_e(k)$ has an extra $i \langle k|\langle \partial_k \chi_m(k)|$ term coming from the $i\partial_k$ differentiating the momentum dependence of the basis ket. The $iU\tilde{\partial}_k U^\dagger$ that appears on the right of Eq. \ref{UXU} can now be understood as canceling off the contribution of this extra term. 

Eq. \ref{kX} now allows us to tackle the question of where $H_{MF}$ comes from. In the spirit of the adiabatic approximation where we forbid transitions between different eigenspinor states, we retain only the diagonal eigenspinor terms of the $iU\partial_k U^\dagger$ piece in Eq. \ref{kX} in the adiabatic approximation to $\hat{x}$ -
\[
	\langle k|\hat{x}_{\text{ad}} \equiv \sum_m \big( |\chi_m(k)\rangle i \langle \chi_m(k)|\partial_k \chi_{m}\rangle\langle k, m| + |\chi_m(k)\rangle i\partial_k\langle  k,m| \big)
\]
(Note that compared to Eq. \ref{kX} we don't sum over the $m'\neq m$ terms in the first piece. )  Substituting in the expansion of the $\sum_m |\chi_m(k)\rangle i\partial_k\langle  k,m|$ from Eq. \ref{cmkm} gives
\begin{equation}
\langle k|\hat{x}_{\text{ad}} = i\langle \partial_k k| \otimes \mathbf{I}_\sigma + \langle k|\otimes \sum_{m,m'\neq m} |\chi_m\rangle i\langle  \partial_k\chi_m|\chi_{m'}\rangle\langle\chi_{m'}|. \label{kxad}
\end{equation}

The first piece on the right $i\langle \partial_k k| \otimes \mathbf{I}_\sigma$  is just $\langle k|\hat{x}$ while the latter piece is exactly the $H_{MF}$ (without the factor of $E$ and with an extra $\langle k|$)  in Eq. \ref{Hmf}. This shows that $H_{MF}$ originates from taking an adiabatic approximation to $\langle k|\hat{x}$. 

The above results can also be obtained from the common usage notation of Eq. \ref{UXU} by splitting up the eigenspinor frame $iU\tilde{\partial}_k U^\dagger = \hat{d} + \hat{o}$ into a \textit{d}iagonal part $\hat{d}$ and \textit{o}ff-diagonal part $\hat{o}$ with
\[
	\hat{d} =  \sum_m |\chi_m\rangle\langle \chi_m|i\partial_k\chi_m\rangle\langle \chi_m|, \hat{o} =  \sum_{m, n\neq m}  |\chi_m\rangle\langle m|i\partial_k \chi_n\rangle\langle \chi_n|
\]
We have, from Eq. \ref{UXU},
\begin{equation}
	U^\dagger ( \hat{x}_e + iU\tilde{\partial}_k U^\dagger) U = \hat{x}_l\Rightarrow U^\dagger ( \hat{x}_e + \hat{d}  ) U = \hat{x}_l - U^\dagger \hat{o} U \label{UoU},
\end{equation}
so that
\begin{equation}
	U^\dagger (\hat{x}_e + \mathrm{diag}(i U\tilde{\partial}_kU^\dagger)) U = \hat{x}_l  -i \sum_{m, n\neq m}  |\chi_m\rangle\langle \chi_m|i\partial_k \chi_n\rangle\langle \chi_n| \label{UoU1}.
\end{equation} 
The left hand side of Eq. \ref{UoU1} has the physical meaning of transforming the adiabatic approximation of $\langle k|\hat{x}$ from the eigenspinor frame back to the laboratory frame. With reference to Eq. \ref{Hmf}, the second term on the right hand side of Eq. \ref{UoU1} $ -i \sum_{m, n\neq m}  |\chi_m\rangle\langle \chi_m|i\partial_k \chi_n\rangle\langle \chi_n|$ is $H_{MF}/E_x$ as expected. 

We therefore conclude this section by summarizing how the results in this and the preceding sections have addressed our motivations in the introduction. 

(i) We raised the issue of how taking the \textit{off}-diagonal elements of $iU\tilde{\partial_k}U^\dagger$ in Eq. \ref{UHU} is consistent with taking an adiabatic approximation.  We now see from Eq. \ref{UoU} that an effective `magnetization' corresponding to the \textit{opposite} direction of the \textit{off-diagonal} terms of $-iU\tilde{\partial}_k U^\dagger$ comes from retaining only the diagonal elements of $iU\tilde{\partial}_k U^\dagger$ in the eigenbasis frame. The key point to note in establishing this equivalence is that we need to transform the eigen-basis frame $\hat{x}_e + \mathrm{diag}(i U\tilde{\partial}_k U^\dagger)$ back to the laboratory frame and not just $\mathrm{diag}(i U\tilde{\partial}_k U^\dagger)$ alone. 

(ii) With the above formalism, we can establish the link between the Kubo formula and the unitary transformation formalism. We found that the Kubo formula can be interpreted as the first order perturbation response to the adiabatic approximation of the position dependence of the electric potential, which we termed as $H_{MF}$. 

\section{Sundaram-Niu wavepacket formalism} 

We next demonstrate that the Heisenberg equations of motion where $E_x \hat{x}$ is replaced by $H_{MF}$ reproduces the dynamics of the position and spin evolution predicted by the Sundaram-Niu wavepacket (SN) formalism \cite{SN0} and the extension of the latter by Cheng and Niu \cite{SN1} to include spin. Physically, the SN formalism revolves around a wavepacket localized about position $x_0$ and momentum $k_0$. This provides an avenue for us to study quantum states which have, in a sense, defined momentum and position values despite the Heisenberg uncertainty principle. Operationally, the position localization offers an elegant solution for studying the effects of perturbations which have the form of $E_x\hat{x}$,  which corresponds to, for example, the potential induced by a uniform DC electric field. In applying the standard first-order time-independent perturbation theory in a SOI system one needs to evaluate matrix elements like $\langle k,m|\hat{x}|k,m\rangle$.  One might naively conclude that such a term might diverge as $x \rightarrow \pm \infty$. The real space localization of the wavepacket around $x_0$ offers the reassurance that we need only take into account the effects of $E\hat{x}$ at the immediate neighbourhood of $x_0$ thus avoiding the divergence at $|x| \rightarrow  \infty$. 

In Cheng and Niu's extension of the SN formalism to include spin, the wavepacket state $|\Psi\rangle$ is given by 
\[
	|\Psi\rangle = \sum_m |k, m\rangle \eta_m a(k) \label{wpPsi}.
\]
Here the $\eta_i$s are the \textit{normalized} weight of the $i$th eigenspinors such that $\sum_i \eta^*_i\eta_i = 1$.  $\eta_i$ is a dynamic variable much like $x$ and $k$ while $a(k)$ is an envelope function peaking around $k=k_0$ where it has the form $\exp(i\gamma(k))$. The requirement that the wavepakcet is localized around $x_0$, $\langle \Psi|\hat{x}|\Psi\rangle = x_0$, imposes some conditions on $\gamma(k)$. 
Relegating the details of the derivation to the appendix, the time evolution of the expectation values of the spin and position operators when $H_0$ does not have explicit position or time dependence is 
\begin{eqnarray}
	\partial_t (\eta^*_i \sigma_{ij} \eta_j) &=& \dot{k}_a (  \eta^*_{l} \langle \chi_{l}|\partial_{k_a}\chi_i\rangle\sigma_{ij}\eta_j - \eta^*_i\sigma_{ij}\langle \chi_j|\partial_{k_a} \chi_l \rangle \eta_l) \label{dtSigma} \\
	\dot{x}_a &=& (\partial_{k_a} e_{\vec{k},m})\eta_m^*\eta_m + i\dot{k_b}\eta^*_m ( \langle \partial_{k_b}\chi_m|\partial_{k_a} \chi_{m'}\rangle  - \langle \partial_{k_a}\chi_m|\partial_{k_b} \chi_{m'}\rangle)\eta_m \label{dtx} 
\end{eqnarray} 
where $\sigma_{ij}\equiv \langle \chi_i(\vec{k})|\sigma|\chi_j(\vec{k})\rangle$. 
\section{Validity of $H_{MF}$. } 
We have seen in the previous section that the Kubo formula can be interpreted as yielding the expectation values of observables under the perturbation of $H_{MF}$. Based on Eq. \ref{Hmf}, this motivated the consideration of 
\begin{equation}
	H_{\mathrm{eff}}(\vec{k}) \equiv H_0(\vec{k}) + E _x\sum_{m, m' \neq m} |\chi_{m'}(k)\rangle\langle i \partial_{k_x} \chi_{m'}(k)|\chi_m(k)\rangle\langle\chi_m(k)|  \label{Heff}
\end{equation}
as the effective Hamiltonian for capturing the effects of the electric field at a specific value of $\vec{k}$ for the original Hamiltonian $H = H_0(\hat{\vec{p}}) + E_x\hat{x}$. We have shown that the expectation values of observables calculated with $H_{\mathrm{eff}}$ will, by construction, match those calculated by the Kubo formula to first order in $E_x$. We now show that applying the Heisenberg equations of motion on Eq. \ref{Heff} will also reproduce the \textit{dynamics} of the position and spin operators as predicted by the SN wavepacket formalism in Eqs. \ref{dtx} and \ref{dtSigma} respectively. 

Now returning to multiple spatial dimensions, the Heisenberg equation of motions are given by (for notational simplicity we temporarily denote $|\chi_m(k)\rangle$ simply as $|m\rangle$)  
\begin{eqnarray*}
	\langle m | \dot{x}_i |m \rangle  &=& ... + \langle m | \left( \sum_{a \neq b}  -i[x_i, |a\rangle\langle a|i \partial_{k_j} b\rangle\langle b|] \right) |m\rangle E_j \\
	&=& + i  (-\langle \partial_{k_i} m|\partial_{k_j} m\rangle + \langle \partial_{k_j} m|\partial_{k_i}m\rangle ) E_j 
\end{eqnarray*}
Noting $\dot{k}_j = -E_j$ with a minus sign, this result agrees with the generalized Sundaram-Niu result of Eq. \ref{dtx}. 

Similarly, if we compare the Heisenberg equation of motion for a spin operator $\sigma = \sum_{ab} |a\rangle\sigma_{ab}\langle b|$ we have 
\begin{eqnarray*}
\langle m | \dot{\sigma} |m \rangle  &=& ... + \langle m | \left( \sum_{a \neq b}  -i[\sigma, |a\rangle\langle a|i \partial_{k_j} b\rangle\langle b|] \right) |m\rangle E_j \\
	&=& ... +  (\sigma_{ma}\langle a|\partial_{k_j}m\rangle -  \langle m|\partial_{k_j}a\rangle\sigma_{am})E_j 
\end{eqnarray*}
which agrees with the generalized SN equation for spin evolution, i.e. Eq. \ref{dtSigma}.   
The agreement of the dynamics of spin and position of $H_{\mathrm{eff}}$ with the SN wavepacket approach provides a physical reason for why it may be justifiable to do away with the operator $\hat{x}$ in Eq. \ref{Heff}. The localization of the wavepacket around $x_0$ (which we may set to 0), implies that the wavepacket only `feels' the influence of position-dependent terms (such as the electric field) at and around the immediate neighbourhood of $x_0$. The irrelevance of the position dependent terms away from $x_0$ makes it unnecessary to include the entire position-dependent potential inside the effective Hamiltonian (recall that $\hat{x} = \int \mathrm{d}x |x\rangle x \langle x|$ is actually the integral of position eigenvalues and position projectors over all sapce) except at $x_0$.  The fact that influence of  $E_x\hat{x}$ away from $x_0=0$ may be ignored, which would be divergent if we had considered all points in space, is perhaps also reflected in the discarding of the divergent $\frac{1}{\omega}, \omega\rightarrow 0$ term in the lead-up to Eq. \ref{kuboX1}.   $H_{MF}$ can then be interpreted as accounting for the effects of the \textit{gradient} of the position dependent potential at $x_0=0$.

In the remainder of this paper, we show that Eq. \ref{Heff} offers three advantages over the direct use of the Kubo formula Eq. \ref{KuboW1} .

i. Eq. \ref{Heff} suggests the possibility of new spin current terms that are not captured by the Kubo formula, as we shall proceed to elaborate on in the next section.  

ii. The usual Kubo formula Eq. \ref{KuboW1} may break down when there are degeneracies because of the $e_a-e_b$ factor in the denominator. Eq. \ref{Heff} validates the use of the degenerate perturbation theory prescription of finding the linear combinations of degenerate eigenstates $|\chi_m\rangle$s for which the $\langle \chi_m | V | \chi_n \rangle = 0$ when $\langle \chi_m |H_0|\chi_m \rangle = \langle \chi_n |H_0|\chi_n\rangle$ for $m \neq n$. 
 
iii. The reasoning behind the construction of Eq. \ref{Heff} suggests a similar approach to constructing an effective Hamiltonian for the effects of a spatially varying magnetization. We describe this in Sect. \ref{sec:varMag}. 

We demonstrate points (i) and (ii) in the next section where we discuss spin currents in the Luttinger heavy hole system, and point (iii) in the section which follows thereafter. 

\section{Murakami-Fujita spin current} 

In the previous section, we have shown that $\langle \chi_m| -i[O, H_{MF}]|\chi_m\rangle$ matches the results of the Sundaram-Niu wavefunction formalism for the operators $O=\hat{x},\vec{\sigma}$. This suggests that in general, there is a contribution to the spin current linear in $E$ of the form 
\begin{equation}
	\langle j^{i(MF)}_{\sigma_\alpha}  \rangle_{\vec{k},m} \equiv \langle \chi_m(\vec{k})| \frac{1}{2}\{ \sigma_\alpha, -i[x_i, H_{MF}]\} |\chi_m(\vec{k})\rangle \label{JiMF}. 
\end{equation} 
This is in addition to the usual term \cite{Sinova} calculated from the Kubo formula
\begin{equation}
	\langle j^{i(\text{Kubo})}_{\sigma_\alpha}  \rangle_{\vec{k},m} = 2\mathrm{Re} \langle \chi_m(\vec{k})|\frac{1}{2} \{ \sigma_\alpha, -i[x_i, H_0] \} |\chi_m^1(\vec{k})\rangle. \label{JiKubo} 
\end{equation}
(As a recap $|\chi_m^1\rangle$ was defined in Eq. \ref{chi1} as the first order correction to $|\chi_m\rangle$ due to the perturbation of $H_{MF}$. )  

This additional spin current contribution has been derived previously for the case of Rashba SOC \cite{SGTSciRep}. We now evaluate this new spin current for the spin 3/2 Luttinger system which was studied previously by Murakami \cite{Murakami}. 

The Hamiltonian for an isotropic spin 3/2 Luttinger system is given by 
\[
	H = \frac{1}{2m} \left(( \gamma_1 + \frac{5}{2} \gamma_2) k^2 \otimes \mathbf{I}_\sigma - 2 \gamma_2 (\vec{k}\cdot\vec{S})(\vec{k}\cdot\vec{S}) \right)
\]
where $\gamma_1$ and $\gamma_2$ are material dependent parameters and $\vec{S}$ are the spin 3/2 operators. 

The eigenstates consist of two pairs of degenerate energy states,  the light hole states $|L1\rangle$ and $|L2\rangle$ with eigenenergy $\frac{k^2}{2m}(\gamma_1 + 2\gamma_2)$ and heavy hole states $|H1\rangle$ and $|H2\rangle$ with eigenenergy $\frac{k^2}{2m}(\gamma_1 - 2\gamma_2)$. The $m$ indices in the notation $|\chi_m\rangle$ which we have been using in the previous sections hence run over $m = \{ L1,L2,H1,H2 \}$.  Murakami used the unitary transformation $U = \exp(i \theta \sigma_y)\exp(i \phi \sigma_z)$ to transform $\hat{x}$ from the lab frame to the eigenspinor frame $\hat{x}_l \rightarrow U\hat{x}U^\dagger = \hat{x}_e + i U\partial_k U^\dagger$. Recalling that $U = \sum_{m, i} |m\rangle\langle m|\lambda_i\rangle \langle \lambda_i|$, we read off from the matrix exponential $U$ that   
\begin{eqnarray}
|H1\rangle &=& \begin{pmatrix}e^{-\frac{3 i \phi }{2}} \cos ^3\left(\frac{\theta }{2}\right) \\ \frac{1}{4} \sqrt{3} e^{-\frac{i \phi }{2}} \sin ^2(\theta ) \csc \left(\frac{\theta }{2}\right) \\ \frac{1}{2} \sqrt{3} e^{\frac{i \phi }{2}} \sin \left(\frac{\theta }{2}\right) \sin (\theta ) \\ e^{\frac{3 i \phi }{2}} \sin ^3\left(\frac{\theta }{2}\right)\end{pmatrix}, |H2\rangle = \begin{pmatrix}-e^{-\frac{3 i \phi }{2}} \sin ^3\left(\frac{\theta }{2}\right) \\ \frac{1}{2} \sqrt{3} e^{-\frac{i \phi }{2}} \sin \left(\frac{\theta }{2}\right) \sin (\theta ) \\ -\frac{1}{4} \sqrt{3} e^{\frac{i \phi }{2}} \sin ^2(\theta ) \csc \left(\frac{\theta }{2}\right) \\ e^{\frac{3 i \phi }{2}} \cos ^3\left(\frac{\theta }{2}\right)\end{pmatrix} \nonumber \\
|L1\rangle &=& \begin{pmatrix}-\frac{1}{4} \sqrt{3} e^{-\frac{3 i \phi }{2}} \sin ^2(\theta ) \csc \left(\frac{\theta }{2}\right) \\ \frac{1}{4} e^{-\frac{i \phi }{2}} \left(\cos \left(\frac{\theta }{2}\right)+3 \cos \left(\frac{3 \theta }{2}\right)\right) \\ -\frac{1}{4} e^{\frac{i \phi }{2}} \left(\sin \left(\frac{\theta }{2}\right)-3 \sin \left(\frac{3 \theta }{2}\right)\right) \\ \frac{1}{2} \sqrt{3} e^{\frac{3 i \phi }{2}} \sin \left(\frac{\theta }{2}\right) \sin (\theta )\end{pmatrix}, |L2\rangle = \begin{pmatrix}\frac{1}{2} \sqrt{3} e^{-\frac{3 i \phi }{2}} \sin \left(\frac{\theta }{2}\right) \sin (\theta ) \\ \frac{1}{4} e^{-\frac{i \phi }{2}} \left(\sin \left(\frac{\theta }{2}\right)-3 \sin \left(\frac{3 \theta }{2}\right)\right) \\ \frac{1}{4} e^{\frac{i \phi }{2}} \left(\cos \left(\frac{\theta }{2}\right)+3 \cos \left(\frac{3 \theta }{2}\right)\right) \\ \frac{1}{4} \sqrt{3} e^{\frac{3 i \phi }{2}} \sin ^2(\theta ) \csc \left(\frac{\theta }{2}\right)\end{pmatrix} \label{LutEigen} 
\end{eqnarray}
where $(k, \theta,\phi)$ are the $k$-space spherical coordinates. 

Following the prescription of Eq. \ref{Hmf},  $H_{MF}$ for an electric field in the $z$ direction reads (in the usual spin 3/2 laboratory spin $z$ basis) 
\[ 
	H_{MF} = i \frac{E_z}{2k} \sin(\theta) \left(
\begin{array}{cccc}
 0 & -\sqrt{3} e^{-i \phi } & 0 & 0 \\
 \sqrt{3} e^{i \phi } & 0 & -2 e^{-i \phi } & 0 \\
 0 & 2 e^{i \phi } & 0 & -\sqrt{3} e^{-i \phi } \\
 0 & 0 & \sqrt{3} e^{i \phi } & 0 \\
\end{array}
\right).
\]  
It can be shown straightforwardly that choosing alternative linear combinations of the degenerate states (i.e. by substituting $|L1'\rangle = |L1\rangle \alpha_1 + |L2\rangle\beta_1, |L2'\rangle = |L1\rangle \alpha_2 + |L2\rangle\beta_2$, expanding and enforcing orthonormality between $|L1'\rangle$ and $|L2'\rangle$, and likewise for the heavy hole states) does not change the final expression for $H_{MF}$.

Defining 
\[
	\langle\langle J^{i(\text{MF})}_{\sigma_\alpha} \rangle\rangle_m \equiv \int^{k_F}_{0} \mathrm{d}k \int \mathrm{d}\theta\mathrm{d}\phi \ \langle \chi_m(\vec{k})| j^{i(MF)}_{\sigma_\alpha}  |\chi_m(\vec{k})\rangle
\]
with $\langle\chi_m| j^{i(MF)}_{\sigma_\alpha}  |\chi_m\rangle$ defined in Eq. \ref{JiMF} gives, after straightforward calculations for an electric field in the $z$ direction, 
\begin{eqnarray*}
	&& \langle\langle J^{x(MF)}_{\sigma_x, \sigma_z }\rangle\rangle_{LH,HH} = \langle\langle J^{y(MF)}_{\sigma_y, \sigma_z }\rangle\rangle_{LH,HH} = 0, \\
	&& \langle\langle J^{z(MF)}_{\sigma_x,\sigma_y, \sigma_z }\rangle\rangle_{LH,HH} = 0, \\
	&& \langle\langle J^{x(MF)}_{\sigma_y}\rangle\rangle_{HH} =  -\langle\langle J^{y(MF)}_{\sigma_x}\rangle\rangle_{HH} = \frac{3e}{\hbar} E_z k_f \pi,  \\
	&& \langle\langle J^{x(MF)}_{\sigma_y}\rangle\rangle_{LH} =  -\langle\langle J^{y(MF)}_{\sigma_x}\rangle\rangle_{LH} = \frac{e}{3\hbar} E_z k_f \pi
\end{eqnarray*} 
where we now write out $e$ and $\hbar$ (which have previously been set to 1) explicitly to show that the results are dimensionally consistent.  
We will now show that the above spin current terms differ from those obtained via Kubo formula $\langle\langle J^{i(\text{Kubo})}_{\sigma_\alpha} \rangle\rangle_m$. As a side-note, in implementing the Kubo formalism, the naive substitution of the eigenstates Eq. \ref{LutEigen} into the Kubo formula Eq. \ref{stdKub1} is problematic because the degeneracy of the eigenstates leads to the presence of terms in which the numerator, proportional to the $H_{MF}$ induced transition between degenerate eigenstates, is finite but the denominator, equal to the energy difference between the degenerate states, is zero. Now that we understand that Eq. \ref{stdKub1} is nothing but standard time-independent perturbation theory applied for the perturbation of $H_{MF}$, we are justified in applying degenerate perturbation theory. Specifically, for each pair of degenerate states say  $|L1\rangle$ and $|L2\rangle$, we diagonalize the projection of $H_{MF}$ in the basis of the degenerate states, in order to find the linear combinations of the degenerate states $|\tilde{L1}\rangle = |L1\rangle\alpha_1 + |L2\rangle\beta_1$,  $|\tilde{L2}\rangle = |L1\rangle\alpha_2 + |L2\rangle\beta_2$ for which $\langle\tilde{L1}|H_{MF}|\tilde{L2}\rangle = 0$ and similarly for the heavy hole states. Using the $|\tilde{L}_i\rangle$ states instead of the original linear combinations in Eq. \ref{LutEigen} ensures that the numerator in the Kubo formula is zero when the denominator is, so that these terms in the Kubo formula can be ignored.

In implementing the above prescription, we find that the $|H1\rangle$ and $|H2\rangle$ states in Eq. \ref{LutEigen} are already diagonal with respect to $H_{MF}$. The linear combination of the light hole states diagonalizing $H_{MF}$ are 
\begin{eqnarray*}
|\tilde{L1}\rangle &=& \begin{pmatrix}-\frac{1}{2} i \sqrt{\frac{3}{2}} e^{\frac{1}{2} i (\theta -3 \phi )} \sin (\theta ) \\ \frac{i \left(1+3 e^{2 i \theta }\right) e^{-\frac{1}{2} i (\theta +\phi )}}{4 \sqrt{2}} \\ \frac{e^{\frac{1}{2} i (\theta +\phi )} (2 \cos (\theta )+i \sin (\theta ))}{2 \sqrt{2}} \\ \frac{1}{2} \sqrt{\frac{3}{2}} e^{\frac{3 i \phi }{2}} \sin \left(\frac{\theta }{2}\right) \sin (\theta ) \left(\cot \left(\frac{\theta }{2}\right)+i\right)\end{pmatrix},
\\ |\tilde{L2}\rangle &=& \begin{pmatrix}\frac{1}{2} i \sqrt{\frac{3}{2}} e^{-\frac{1}{2} i (\theta +3 \phi )} \sin (\theta ) \\ \frac{e^{-\frac{1}{2} i (\theta +\phi )} (-\sin (\theta )-2 i \cos (\theta ))}{2 \sqrt{2}} \\ \frac{\left(3+e^{2 i \theta }\right) e^{-\frac{1}{2} i (3 \theta -\phi )}}{4 \sqrt{2}} \\ \frac{1}{2} \sqrt{\frac{3}{2}} e^{\frac{3 i \phi }{2}} \sin \left(\frac{\theta }{2}\right) \sin (\theta ) \left(\cot \left(\frac{\theta }{2}\right)-i\right)\end{pmatrix}.
\end{eqnarray*}

We then find, after straightforward but tedious calculations, that the only non-zero $\langle\langle j^{i(\text{Kubo}}_{\sigma\alpha}\rangle\rangle_m$s are 
\[
	\langle\langle j^{x(\text{Kubo)}}_{\sigma_y}\rangle\rangle_{HH} = \langle\langle j^{y(\text{Kubo)}}_{\sigma_x}\rangle\rangle_{LL} =  -\langle\langle j^{y(\text{Kubo)}}_{\sigma_x}\rangle\rangle_{HH} = - \langle\langle j^{x(\text{Kubo)}}_{\sigma_y}\rangle\rangle_{LL} = \frac{e E_z k_f}{2\hbar\gamma_2}\pi(\gamma_1+2\gamma_2).
\]

\section{Position-dependent degree magnetization}
\label{sec:varMag}
We next move on to study the effects of a weak position dependence of the coupling to the internal degrees of freedom in the form $M(\hat{x})_i\kappa_i$ which we shall collectively refer to as the `magnetization' (in analogy with $M_i\sigma_i$ for $\sigma_i$ being the real spin) for short. We assume that the position dependence is weak enough for a first order gradient expansion of the magnetization to suffice -- 
\[
	H = B(\vec{k})_i\kappa_i + M_i(0) \kappa_i + (\partial_x M_i)\kappa_i \hat{x}.
\]

The $B(\vec{k})_i\kappa_i$ is the momentum-dependent SOC term and the  $M_i(0)$ are the magnetizations at the fixed $x$=0 and are position-independent. We take $H_0 = (B(\vec{k})_i + M_i(0))\kappa_i$, and treat $(\partial_x M_i)\kappa_i \hat{x}$ as the perturbation.  Note that this is a different, but not necessarily mutually exclusive, limit from the more often studied case (Ref. \onlinecite{Grp4} and references therein) where the SOC is treated as a perturbation to the position dependent magnetization under the strong magnetization limit. 

Analogous to what we have done in the previous section, we now seek an expression for 
\begin{equation}
	\langle k|\hat{x}\kappa_i = \sum_{m} \left( \langle k,m|\hat{x}\kappa_i \sum_{m'} \int \mathrm{d}k' |k',m'\rangle\langle k',m'| \right)
\end{equation}
where the $|k,m\rangle$s are the eigenstates of $H_0$. 

Employing a similar series of steps in our earlier derivation of $\langle k,m|\hat{x}|k',m'\rangle$, it can be shown that (for brevity, we have dropped the subscript $i$ in $\kappa$) 
\begin{eqnarray}
	&&\langle k,m|\hat{x}\kappa|k',m'\rangle \nonumber \\
	 &=& i (\partial_k (\delta(k'-k)\langle \chi_m(k)|\kappa|\chi_{m'}(k')\rangle) - \delta(k'-k)\langle i\partial_k \chi_m(k) |\kappa|\chi_{m'}(k')\rangle \label{Xsdk1} \\
	&=& -i \partial_{k'} (\delta(k'-k)\langle \chi_m(k)|\kappa|\chi_{m'}(k')\rangle) + \delta(k'-k)\langle \chi_m(k) |\kappa|i\partial_{k'}\chi_{m'}(k')\rangle \label{Xdsk2} 
\end{eqnarray}

Using the expansion of  Eq. \ref{Xsdk1}  for $\langle k,m|\hat{x}\kappa|k',m'\rangle$ gives
\begin{equation}
\int \mathrm{d}k' |k,m\rangle\langle k,m|\hat{x}\kappa|k',m'\rangle\langle k',m'|=   |k,m\rangle    (\langle \chi_m|\kappa|i \partial_k \chi_{m'}\rangle\langle k,m'| + \langle \chi_m|\kappa|\chi_{m'}\rangle i\partial_k \langle k,m'|).  \label{kmsxkm1}
\end{equation}
Notice that each of the two terms on the right hand side are not individually Hermitian by themselves.  The remedy to this is to take half of the right hand side of Eq. \ref{kmsxkm1}, and add this to half of the left hand side of Eq. \ref{kmsxkm1} evaluated using the expansion of Eq. \ref{Xdsk2}.  The latter is (half of) the Hermitian conjugate of Eq. \ref{kmsxkm1}.   The symmeterized version of Eq. \ref{kmsxkm1} thus reads
\begin{eqnarray}
&& \int \mathrm{d}k' |k,m\rangle\langle k,m|\hat{x}\kappa|k',m'\rangle\langle k',m'| \nonumber \\
&=&\frac{1}{2} \Big( (|k,m\rangle    (\langle \chi_m|\kappa|i \partial_k \chi_{m'}\rangle - i \langle \partial_k \chi_m|\kappa|\chi_{m'}\rangle) \langle k,m'| +  \nonumber \\
&& (-|i\partial_k  k,m\rangle\langle \chi_m|\kappa|\chi_{m'}\rangle \langle k,m'| + |k,m\rangle \langle \chi_m|\kappa|\chi_{m'}\rangle \langle i\partial_k k,m'|)  \Big) \label{kmsxkm2}
\end{eqnarray} 

We now take the adiabatic approximation of retaining only the diagonal terms in the eigenbasis.  Differing from our earlier experience with the expansion of $\int \mathrm{d}k' |k,m\rangle\langle k,m|\hat{x}|k',m'\rangle\langle k',m'|$ where the terms that come with the $\langle i\partial_k|$ and $|i\partial_k \rangle$ are already diagonal in the eigenspinor basis, the corresponding two terms here in the last line of Eq. \ref{kmsxkm2} do contain off-diagonal elements.  This raises the question of whether we should retain the $m' \neq m$ elements in these terms. 

If we retain $m \neq m'$ terms in the terms containing $|i\partial_k k,m\rangle$ and its c.c. in the last line of Eq. \ref{kmsxkm2} , we have 

\begin{eqnarray*}
	&& \sum_m |k,m\rangle   \Big(  \langle \chi_m|\kappa|i \partial_k \chi_m\rangle\langle k,m| + \sum_{m'} \langle \chi_m|\kappa|\chi_{m'}\rangle i\partial_k \langle k,m'|)  + \text{h.c.} \Big) \\
	&=& (|k\rangle i\kappa \langle \partial_k k| - \sum_{m \neq m'} |k,m\rangle  \langle \chi_m|\kappa|i\partial_k \chi_{m'}\rangle\langle k,m'|)  + \text{h.c.}
\end{eqnarray*}

We treat the second term and its Hermitian conjugate as the perturbation 
\[
	\tilde{H}_{MF;\text{mag}} = \frac{1}{2}(\partial_x M_i)  \sum_{m \neq m'}  ( |k,m\rangle  \langle \chi_m|\kappa|i\partial_k \chi_{m'}\rangle\langle k,m'| + \text{h.c.} ). 
\]
Comparing this against Eq. \ref{Hmf} for an electric field, one can interpret $(\partial_{x_i} M)\kappa_i\hat{x}$ as a spin dependent `electric field'. One could, however, obtain a closer agreement with the predictions of the Sundaram-Niu wavepacket formalism by discarding the $m \neq m'$ terms in the last line of Eq. \ref{kmsxkm2} in keeping with the spirit of the adiabatic approximation of forbidding transitions between different eigenstates. We obtain

\begin{equation}
\frac{1}{2} (\langle k|(\hat{x}\kappa_i)_{\mathrm{ad}} + \mathrm{h.c.}) = \frac{1}{2} \sum_m |k,m\rangle \big(    \langle \chi_m |\kappa|i \partial_k \chi_m\rangle\langle k,m| + \langle \chi_m|\kappa|\chi_{m}\rangle (\langle i \partial_k \chi_{m}|\langle k |   +  \langle \chi_m | \langle i\partial_k|) + \mathrm{h.c.} \big).  \label{adMx1}
\end{equation}
 
 This is the analogue of Eq. \ref{kxad}, symmeterized to ensure that the term remains Hermitian. The $\frac{1}{2}  |\chi_m\rangle \langle \chi_m| \kappa_i|\chi_m\rangle\langle\chi_m|\otimes (|k\rangle\langle i\partial_k| - |i\partial_k\rangle\langle k| )$ term in Eq. \ref{adMx1} (the last term in Eq. \ref{adMx1} just to the left of the `h.c.' and its conjugate) ) is analogous to $i \langle \partial_k k|\otimes\mathbf{I}_\kappa$ in Eq. \ref{kxad}. This analogy is in the sense that $\frac{1}{2}  |\chi_m\rangle \langle \chi_m| \kappa_i|\chi_m\rangle\langle\chi_m|\otimes (|k\rangle\langle i\partial_k| - |i\partial_k\rangle\langle k| )$ can be interpreted as the projection of the spin part of $\hat{x}\kappa$ into the eigenspinor basis of $H_0$, just like $i \langle \partial_k k|\otimes\mathbf{I}_\kappa$ is actually $(\sum_m (|\chi_m\rangle\langle\chi_m |\mathbf{I}|\chi_m\rangle\langle\chi_m|)\otimes i \langle \partial_k k|$. We treat the remaining terms in Eq. \ref{adMx1},
\begin{equation}
H_{MF;\text{mag}} = \frac{\partial_x M}{2}|k,m\rangle( \langle \chi_m |\kappa|i \partial_k \chi_m\rangle\langle k,m| + \langle \chi_m|\kappa|\chi_{m}\rangle \langle i \partial_k \chi_{m}|\langle k | + \mathrm{h.c.}). \label{HmfMag}
\end{equation}
as perturbations to $|k,m\rangle$ and calculate the  perturbed states $|k,m^1\rangle_{\text{mag}} = \sum_{n\neq m} \frac{|k,n\rangle \langle\chi_n|H_{MF;\text{mag}}|\chi_m\rangle}{e_{n,k}-e_{m,k}}$. The first order correction to the velocity due to $H_{MF;\text{mag}}$,  $2\mathrm{Re} (\langle k,m|-i[x, H_0]|k,m^1\rangle_{\text{mag}} )$ then gives the SN velocity Eq. \ref{dtx} with $\dot{k}_{|k,m\rangle} = -(\partial_x M_i) \langle k,m|\kappa_i|k,m\rangle$. This is consistent with treating $(\partial_x M_i)\kappa_i \hat{x}$ as a spin-dependent electric field. 

Besides giving a velocity contribution, the two terms containing an outer ket-bra pair of the form $|\chi_m\rangle\langle \partial_k \chi_m|$ and its h.c. in Eq. \ref{HmfMag} may also give an effective `magnetization' pointing in a different direction from the magnetization direction of $H_0$. The Hamiltonian for this effective magnetization, $H_M$ is given by the off-diagonal terms of $H_{MF;\text{mag}}$, which consist of the second term of Eq. \ref{HmfMag}, $\frac{\partial_x M}{2} |k,m\rangle(\langle \chi_m|\kappa|\chi_{m}\rangle \langle i \partial_k \chi_{m}|\langle k | + \mathrm{h.c.}$. with their spin diagonal components, i.e. the projection of $|\chi_m\rangle\langle \partial_k \chi_m|$ onto $|\chi_m\rangle\langle\chi_m|$  subtracted away. We obtain

 \begin{eqnarray}
 	H_M &=& \frac{1}{2} (\partial_x M_i) \sum_m \langle \chi_m|\kappa_i|\chi_m\rangle ( |\chi_m\rangle\langle i \partial_k \chi_m|  - |\chi_m\rangle\langle i\partial_k \chi_m|\chi_m\rangle\langle\chi_m| + \mathrm{h.c.} ) \nonumber \\
	&=& \frac{1}{2}  (\partial_x M_i) \sum_{m, n\neq m}  (|\chi_m\rangle \langle \chi_m|\kappa_i|\chi_m\rangle \langle i\partial_k \chi_m|\chi_n\rangle\langle\chi_n| + \mathrm{h.c.} ) \label{Mom}
 \end{eqnarray}

For a two-level system where the $\kappa_i$s correspond to the Pauli matrices, the effective magnetization evaluates to zero. However, in systems with multiple coupled internal degrees of freedom, such as the TI thin film, described by Eq. \ref{TIham1}, some physically relevant choices of the $\kappa_i$ do result in finite effective magnetizations which in turn result in spin accumulations. 

We illustrate this by numerically evaluating the spin accumulations which result from a $x$ direction $M_i$ gradient for the $M_i$s coupling to $\sigma_a$ and $\sigma_a\tau_z$. The $(\partial_x M_a)\sigma_a$s  correspond to $x$ gradients in $a$ direction magnetization exchange couplings to the top and bottom thin film surfaces where the couplings on both surfaces have the same sign and magnitude. The  $(\partial_x M_{a;z})\sigma_a\tau_z$  correspond to exchange couplings of the same magnitude but opposite signs on the top and bottom surfaces. An exchange coupling with different magnitudes on the top and bottom surfaces (for example, due to differing distances from the ferromagnetic layer on top of the TI thin film) is then given by $(M_a + \alpha M_{a;z}\tau_z)\sigma_a$ with $\alpha$ denoting the degree of asymmetry between the two surfaces. We integrate the spin accumulations in the $\sigma_a\tau_b$ ($\tau_b = (\mathbf{I}_\tau, \tau_z)$) resulting from the magnetization gradient integrated over the Fermi surface, and denote the integrated spin accumulation as $\langle\langle\delta\sigma_a\tau_b\rangle\rangle$. 

Tables I summarizes which combinations of the $(\partial_ x M_{c,d})\sigma_c\tau_d$ gradients result in finite $\langle\langle\delta\sigma_a\tau_b\rangle\rangle$ accumulations when only the $x$, $y$ and $z$ component of the constant $M_i\sigma_i$ magnetization in Eq. \ref{TIham1} has non-zero components.  These correspond to the three scenarios where the magnetization gradient in the $x$ direction is parallel to the constant magnetization, perpendicular to the constant magnetization and in the in-plane direction, and in the out-of-plane directions respectively. A detailed analysis of the spin accumulations resulting from the magnetization gradients will be presented elsewhere. 

\begin{table}
\begin{tabular}{|c|c|c|c|c|}  \hline 
\multicolumn{5}{|c|} {Only $M_x(0)$ non-zero} \\ \hline
Spin accumulation / Magnetization gradient &$\sigma_x $&$\sigma_y $&$\sigma_x \tau_z$&$\sigma_y \tau_z$\\ \hline 
$\langle\langle\delta\sigma_z\rangle\rangle$ &\checkmark & &\checkmark &\checkmark \\ \hline 
$\langle\langle\delta\sigma_z\tau_z\rangle\rangle$ &\checkmark &\checkmark &\checkmark &\checkmark \\ \hline 
\end{tabular} 

\begin{tabular}{|c|c|c|}  \hline 
\multicolumn{3}{|c|} {Only $M_y(0)$ non-zero} \\ \hline
Spin accumulation / Magnetization gradient &$\sigma_x $&$\sigma_x \tau_z$\\ \hline 
$\langle\langle\delta\sigma_z\rangle\rangle$ &\checkmark &\checkmark \\ \hline 
$\langle\langle\delta\sigma_z\tau_z\rangle\rangle$ &\checkmark &\checkmark \\ \hline 
\end{tabular} 

\begin{tabular}{|c|c|c|c|c|c|c|}  \hline 
\multicolumn{7}{|c|} {Only $M_z(0)$ non-zero} \\ \hline
Spin accumulation / Magnetization gradient &$\sigma_x $&$\sigma_y $&$\sigma_z $&$\sigma_x \tau_z$&$\sigma_y \tau_z$&$\sigma_z \tau_z$\\ \hline 
$\langle\langle\delta\sigma_x\rangle\rangle$ & &  &\checkmark &  &\checkmark &\checkmark  \\ \hline 
$\langle\langle\delta\sigma_z\rangle\rangle$ &\checkmark & & &\checkmark & &  \\ \hline 
$\langle\langle\delta\sigma_x\tau_z\rangle\rangle$ & &\checkmark &\checkmark & &\checkmark &\checkmark  \\ \hline 
$\langle\langle\delta\sigma_z\tau_z\rangle\rangle$ &\checkmark & & &\checkmark & &  \\ \hline 
\end{tabular} 

\caption{A checkmark in the cell indicates that a finite spin $\sigma_a\tau_b$ accumulation (rows) exists after integrating over the Fermi surface for a $(\partial_x M_{\sigma_c\tau_d}) \sigma_c\tau_d$ gradient (columns) when only the stated component of the constant $\vec{M}\cdot\vec{\sigma}$ magnetization is present. } 
\end{table}

\section{Conclusion}  
In this work, we showed that the non-equilibrium expectation value of an observable $O$ due to a DC electric field given by the Kubo formula Eq. \ref{Kubo} can be interpreted as the standard time-independent perturbation theory response of the observable $O$ to the perturbation $H_{MF}$. $H_{MF}$ originates from taking the adiabatic approximation to $\langle k|\hat{x}$ in which transitions between different eigenspinor states are forbidden.  We then argued that to first order in the electric field, the effects of $E_x\hat{x}$ on the spin accumulation and currents can be captured in an effective Hamiltonian where $E\hat{x}$ is replaced by $H_{MF}$. We supported this claim by showing that the Heisenberg equations of motion with this effective Hamiltonian reproduces the dynamics for spin and current predicted by the Sundaram-Niu wavepacket formalism. The effective Hamiltonian suggests the existence of a new spin current term which is not captured by the standard Kubo formula, and which we then evaluated for the Luttinger heavy hole system.  Finally, using the same approach used to obtain $H_{MF}$ for an electric field, we derived a corresponding expression for $H_{MF;\text{mag}}$ resulting from a spatially varying magnetization. We showed that $H_{MF;\text{mag}}$, similar to $H_{MF}$, leads to an anomalous velocity as well as an effective magnetization.

\section{Acknowledgments} 
We thank the MOE Tier II grant MOE2013-T2-2-125 (NUS Grant No. R-263-000-B10-112), and the National Research Foundation of Singapore under the CRP Program ``Next Generation Spin Torque Memories: From Fundamental Physics to Applications'' NRF-CRP9-2013-01 (R-263-000-B30-281) and ``Non-Volatile Magnetic Logic and Memory Integrated Circuit Device'' NRF-CRP9-2011-01 (R-263-000-A73-592) for financial support.

\section{Appendix}
\subsection{Derivation of $\langle k|\hat{x}$}

In order to obtain the expression for $\langle k|\otimes\mathbf{I}_\sigma \hat{x}$ in terms of a momentum dependent spin basis we first derive the expression for $\langle k,m|\hat{x}|k',m'\rangle$ -- 

\begin{eqnarray}
	&& \langle k, m| \hat{x} | k', m'\rangle \nonumber \\ 
	&=& \int \mathrm{d}x \langle k,m|x\rangle x \langle x|k',m'\rangle \nonumber \\ 
	&=&  \frac{1}{2\pi} \int \ \mathrm{d}x\ \langle\chi_m(k)|\chi_{m'}(k')\rangle \exp( i (k'-k)x)x  \nonumber  \\ 
	&=& i \big(-\delta_{m,m'} \partial_{k'}   +   \langle \chi_m(k)|\partial_k \chi_{m'}\rangle \big)\delta(k'-k) \nonumber \label{kmXkm}
\end{eqnarray}

With this expression, we can then derive 
\begin{eqnarray*}
	&& \langle k|\hat{x} \nonumber \\
	&=& \ \sum_{m,m'} |\chi_m(k)\rangle \int\mathrm{d}k'\    i \big(-\delta_{m,m'} \partial_{k'}   +   \langle \chi_m(k)|\partial_k \chi_{m'}\rangle \big)\delta(k'-k) \langle k',m'|  \\
	&=&    i \sum_{m,m'}|\chi_m(k)\rangle\langle \chi_m(k)|\partial_k \chi_{m'}\rangle\langle k, m'| +\sum_m |\chi_m(k)\rangle i\partial_k\langle  k,m|  
\end{eqnarray*}

\subsection{SN formalism with spin} 
(We note that the following derivations is not our original work and follow that in Refs. \cite{SN0} and \cite{SN1} with a minor extension. The derivations here are simpler as we do not consider explicit time and position dependence in $H_0$. We have chosen to go into more detail in some portions which have not been described as explicitly in the two references. ) 

Consider the wavepacket state $|\Psi\rangle$ of Eq. \ref{wpPsi}. In order to enforce $\langle \Psi|\hat{x}|\Psi\rangle = x_c$ we first evaluate 
\begin{eqnarray*}
 	 \langle \Psi| x | \Psi \rangle 	&=& \int \mathrm{d}k\ \mathrm{d}k'\ a^*(k)\eta_m^*  \langle k,m | x | k', m'\rangle  \eta_{m'} \\
	&=& i \eta^*_m   \Big(  (-i \partial_k \gamma)\delta_{mm'}    +   \langle \chi_m|\partial_k \chi_{m'}\rangle \Big)\eta_{m'} .
\end{eqnarray*}
from which we conclude that 
\begin{equation}
 	\partial_k \gamma (\eta^*_m\eta_m) = x_c - i\eta^*_m\langle \chi_m|\partial_k \chi_{m'}\rangle\eta_m, \label{dkGamma} 
\end{equation}
a fact which we take note of for now and save for further use later. 

We now proceed to construct the Lagrangian given by 
\[ 
	L = \langle \Psi| i \mathrm{d}_t - (H_0 + H') |\Psi\rangle.
\]

We thus need 
\[
	i \mathrm{d}_t |\Psi\rangle = \int \mathrm{d}k'  \big( |k',m'\rangle \big(  \partial_t\gamma +  \dot{x}_c \partial_{x_c} \gamma) \eta_{m'}  + i \dot{\eta} _{m'} \big)  a(k').
\]
so that
\begin{eqnarray*}
	\langle \Psi |i\mathrm{d}_t \Psi\rangle &=& i \eta_m^*\dot{\eta}_m + (\partial_t\gamma)\eta^*_m\eta_m +  \eta_m^* \dot{x}_c  \langle \chi_m|\partial_{x_c} \chi_{m'}\rangle \eta_{m'}\\
	&=& i \eta_m^*\dot{\eta}_m  + d_t(\gamma  - kx_c) + k\dot{x}_c +  i(\dot{k}\langle\chi_m|\partial_k \chi_{m'}\rangle)  + \dot{x}_c \langle \chi_m|\partial_{x_c}\chi_{m'}\rangle) \eta^*_m\eta_{m'}
\end{eqnarray*}

Now for our purposes the perturbations we have in mind inside $H'$ are proportional to $(\hat{x}-x_c)$ (or position and momentum independent, if we choose to include these as perturbations) . Since by definition $\langle \Psi| \hat{x} |\Psi \rangle = x_c$ then $\langle \Psi | \hat{x}-x_c |\Psi\rangle = 0$, the terms in  $H'$ propotional to $(x-x_c)$ do not contribute to $L$. If we do choose to include terms dependent only on (the parameter) $x_c$ into the perturbation $V(x_c)\sigma$ then $\langle \Psi | H' |\Psi \rangle = \eta^*_m \langle \chi_m|V(x_c)\sigma|\chi_{m'}\rangle \eta_{m'}$. 

For the remainder of this section we consider $|\chi_m\rangle$ to be $\vec{r}$ independent. 
The Lagrangian thus reads, after discarding the total time derivative $\mathrm{d}_t(\gamma  - kx_c)$, 
\begin{equation}
	L = i \eta_m^*\dot{\eta}_m + k\dot{x}_c +  i(\dot{k}\langle\chi_m|\partial_k \chi_{m'}\rangle) - e_{k,m}\eta_m^*\eta_m.
\end{equation}

Restoring multiple spatial dimensions and dropping the $c$ subscript from $x$, the Lagrangian reads
\begin{equation}
	L = i \eta_m^*\dot{\eta}_m + k_a\dot{x}_a +  i(\dot{k_a}\langle\chi_m|\partial_{k_a} \chi_{m'}\rangle)  - e_{\vec{k},m}\eta_m^*\eta_m.
\end{equation}
Taking functional derivatives with respect to $k_a$, we have
\begin{eqnarray}
	\partial_{k_a} L &=& \dot{x}_a + i\eta_m^* \big(\dot{k_b} \partial_{k_a} \langle\chi_m|\partial_{k_b} \chi_{m'}\rangle \big)\eta_{m'} - (\partial_{k_a} e_{\vec{k},m})\eta_m^*\eta_m \label{dkaL}  \\
	\partial_t \partial_{\dot{k_a}} &=& \partial_t ( i \eta^*_m \langle\chi_m|\partial_{k_a} \chi_{m'}\rangle \eta_{m'}  )  \nonumber \\
	&=&  i  \eta^*_m  (\dot{k_b}(\partial_{k_b}  \langle\chi_m|\partial_{k_a} \chi_{m'}\rangle)  ) \eta_{m'}  \label{ddtkaL} 
\end{eqnarray}
so that we end up with the Euler-Lagrange equation
\begin{eqnarray}
	\dot{x}_a &=& (\partial_{k_a} e_{\vec{k},m})\eta_m^*\eta_m + i \dot{k_b}\eta^*_m (\partial_{k_b}  \langle\chi_m|\partial_{k_a} \chi_{m'}\rangle  - \partial_{k_a}\langle\chi_m|\partial_{k_b} \chi_{m'}\rangle)\eta_m  \nonumber \\ 
	&=& (\partial_{k_a} e_{\vec{k},m})\eta_m^*\eta_m + i\eta^*_m (\dot{k_b} ( \langle \partial_{k_b}\chi_m|\partial_{k_a} \chi_{m'}\rangle  - \langle \partial_{k_a}\chi_m|\partial_{k_b} m'\rangle)\eta_m 
\end{eqnarray}

Taking the functional derivative with respect to $\eta_m$ gives 
\[
	\dot{\eta_m^*} = \eta_{m'}^*\big( (\dot{k_a}\langle\chi_{m'}|\partial_{k_a} \chi_{m}\rangle  + \dot{x}_a \langle \chi_{m'}|\partial_{x_a}\chi_{m}\rangle) +i  \langle \chi_{m'}|V(\hat{r})\sigma|\chi_{m}\rangle \big) +  i\eta^*_m e_{\vec{k},m})
\]
so that on taking CC we have 
\[
	\dot{\eta_m} = -\big(\dot{k_a}\langle\chi_{m}|\partial_{k_a} \chi_{m'}\rangle  + \dot{x}_a \langle \chi_{m}|\partial_{x_a}\chi_{m'}\rangle + i  \langle \chi_{m}|V(\hat{r})\sigma|\chi_{m'}\rangle   \big) \eta_{m'} - ie_{\vec{k},m}(\vec{r})\eta_m
\]

Together, these give (concentrating on the $\dot{k}_a$ piece), 
\begin{eqnarray}
	\partial_t \eta^*_i \sigma_{ij} \eta_j &=& \dot{k}_a (  \eta^*_{l} \langle \chi_{l}|\partial_{k_a}\chi_i\rangle)\sigma_{ij}\eta_j - \eta^*_i\sigma_{ij}\langle \chi_j|\partial_{k_a} \chi_l \rangle \eta_l + ... \nonumber \\
	&=& \dot{k}_a (  \eta^*_{l} \langle \chi_{l}|\partial_{k_a}\chi_i\rangle)\sigma_{ij}\eta_j - \eta^*_j\sigma_{ji}\langle \chi_i|\partial_{k_a} \chi_l \rangle \eta_l + ... \text{ (Swap $i\leftrightarrow j$ in second piece)}
\end{eqnarray}

\end{document}